\documentstyle[epsfig]{cupconf}

\ifoldfss
\else
  \ifnfssone
    \newmathalphabet{\mathit}
      \addtoversion{normal}{\mathit}{cmr}{m}{it}
      \addtoversion{bold}{\mathit}{cmr}{bx}{it}
    \newmathalphabet{\mathcal}
      \addtoversion{normal}{\mathcal}{cmsy}{m}{n}
    \else
    \ifnfsstwo
    \fi
  \fi
\fi

\def\GeV{\,{\rm GeV}}
\def\TeV{\,{\rm TeV}}

\def\sec{\,{\rm sec}}
\def\Gyr{\,{\rm Gyr}}

\def\rcm{\,{\rm cm}}
\def\km{\,{\rm km}}

\def\Mpc{\,{\rm Mpc}}

\def\eV{{\,\rm eV}}

\def\cmm2{{\,\rm cm^{-2}}}
\def\cm2{{\,{\rm cm}^2}}
\def\cmm3{{\,{\rm cm}^{-3}}}
\def\gcmm3{{\,{\rm g\,cm^{-3}}}}
\def\kms{\,{\rm km\,s^{-1}}}

\def\mpl{{m_{\rm Pl}}}

\def\la{\mathrel{\mathpalette\fun <}}
\def\ga{\mathrel{\mathpalette\fun >}}
\def\fun#1#2{\lower3.6pt\vbox{\baselineskip0pt\lineskip.9pt
  \ialign{$\mathsurround=0pt#1\hfil##\hfil$\crcr#2\crcr\sim\crcr}}}

\makeatletter
\ifx\CUP@mtlplain@loaded\undefined
\else
\fi
\makeatother
 \makeatletter
 \ifx\CUP@mtlplain@loaded\undefined
   \font\tenbmi=cmmib10 at 10pt
   \font\sevenbmi=cmmib10 at 7pt
   \font\fivebmi=cmmib10 at 5pt

   \newfam\bmifam
   \textfont\bmifam=\tenbmi
   \scriptfont\bmifam=\sevenbmi
   \scriptscriptfont\bmifam=\fivebmi
   
 \fi
 \makeatother
\ifnfsstwo

\fi
\ifnfssone

\fi
\ifoldfss

\fi

\mathchardef\varLambda="0103

\makeatletter
\ifx\CUP@mtlplain@loaded\undefined
\else
\fi
\makeatother
\makeatletter
\ifx\CUP@mtlplain@loaded\undefined
  \font\tenbms=cmbsy10
  \font\sevenbms=cmbsy10 at 7pt
  \font\fivebms=cmbsy10 at 5pt
  \newfam\bmsfam
  \textfont\bmsfam=\tenbms
  \scriptfont\bmsfam=\sevenbms
  \scriptscriptfont\bmsfam=\fivebms
\fi
\makeatother
\title[Inflationary Cosmology]{Inflationary Cosmology}

\author[Michael S. Turner]%
{M\ls I\ls C\ls H\ls A\ls E\ls L\ns S.\ns T\ls U\ls R\ls N\ls E\ls R$^{1,2}$
\thanks{This work was supported by the US DoE (at Chicago and
Fermilab) and the US NASA (at Fermilab).}\ns}

\affiliation{$^1$Departments of Physics and of Astronomy and
Astrophysics, Enrico Fermi Institute, The University of Chicago,
Chicago, IL~~60637-1433, USA\\[\affilskip]
$^2$NASA/Fermilab Astrophysics Center, Fermilab, Batavia, IL~~60510-0500, USA}

\begin{document}
\ifnfssone
\else
  \ifnfsstwo
  \else
    \ifoldfss
      \let\mathcal\cal
      \let\mathrm\rm
      \let\mathsf\sf
    \fi
  \fi
\fi

\maketitle

\begin{abstract}
Inflation is a bold and expansive extension of the Standard Cosmology.
It holds the promise to extend our understanding of the Universe to
within $10^{-32}\sec$ of the big bang and answer most of the pressing
questions in cosmology.  Its boldest assertion is that all
the structure observed in the Universe today arose from quantum-mechanical
fluctuations on subatomic scales.  A flood of cosmological
observations and laboratory experiments are testing inflation and within
five years we should know whether inflation is to become part of a
new standard cosmology.
\end{abstract}

\firstsection 

\section{From Hubble to Zel'dovich \& Novikov}

The value of the Hubble constant has changed by about a factor of ten since
Edwin Hubble's pioneering measurements.  The context in which we
view the Universe has changed even more profoundly.  Until 1964
cosmology was mostly concerned with cosmography; the spirit of
this period was perhaps best captured by Sandage, ``the quest
for two numbers ($H_0$ and $q_0$).''  The discovery of the Cosmic
Background Radiation (CBR) led to the establishment of a physical foundation
for the expanding Universe -- the hot big-bang cosmology.
The 1970s saw this model become firmly established as the Standard Cosmology.
Igor Novikov played an important role in this enterprise,
and with Zel'dovich wrote one of the first reviews of the Standard
Cosmology (\cite{araa1}) as well as a comprehensive and authoritative summary
of it in Volume 2 of {\it Relativistic Astrophysics (Structure and
Evolution of the Universe)} (\cite{relastro}).

The hot big-bang cosmology is a remarkable achievement.  It
provides a reliable account of the Universe from about $10^{-2}\sec$
to the present (\cite{bigbang1a}; \cite{bigbang1b}).  Further, the hot big-bang
model together with modern ideas in
particle physics --- the Standard Model, supersymmetry, grand
unification, and superstring theory --- provides a framework
for sensible speculation all the way back to the Planck epoch,
and perhaps even earlier.  Here too, Novikov was one of the
pioneering figures in the study of the earliest moments (\cite{araa2};
\cite{relastro}).

Building upon the Standard Cosmology, in the 1980s cosmologists and particle
physicists began an ambitious program to root the hot big-bang
model in fundamental physics and to address
a deeper set of questions.  What is the
nature of the ubiquitous dark matter that is the dominant component
of the mass density?  Why does the Universe contain only matter?
What is the origin of the tiny inhomogeneities
that seeded the formation of structure, and how did that structure
evolve?  Why is the portion of the Universe that we can see so
flat and smooth?  What is the value of the cosmological
constant?  What is the origin of the entropy (photons in
the CBR)?  How did the expansion begin---or was there a beginning?
What was the big bang?

In the past fifteen years much progress has been made, and
many believe that the answers to all these questions involve
events that took place during the earliest moments and involved physics
beyond the Standard Model (\cite{eu}).  For example, the matter-antimatter
asymmetry,
quantified as a net baryon number of about $10^{-10}$ per photon, is
believed to have developed through interactions that do not conserve
baryon number or $C$, $CP$ and
occurred out of thermal equilibrium.  If ``baryogenesis'' involved
unification-scale physics the baryon asymmetry developed
around $10^{-34}\sec$ (\cite{kt}); on the other hand, baryogenesis
may have occurred at the weak scale ($T\sim 300\GeV$ and
$t\sim 10^{-11}\sec$) through baryon-number violation within the
Standard Model and $C$, $CP$ violation from physics
beyond the Standard Model (Cohen et al. 1992; \cite{ewbaryo2}).

The most optimistic early-Universe cosmologists (of which I am one)
believe that we are on the verge of solving almost all of the above problems
and extending our knowledge of the Universe back to around $10^{-32}
\sec$ after ``the bang.''  The key is inflation.  Among other
things, inflation has led to the cold dark matter model
of structure formation, whose basic tenets are
scale-invariant density perturbations and
dark matter whose primary composition is slowly moving elementary
particles (e.g., axions or neutralinos).   Cold dark matter
provides a crucial test of inflation and a possible window to physics
at unification-scale energies.   If cold dark matter is
correct, it would complete the Standard Cosmology
by connecting the theorist's smooth early Universe
to the astronomer's Universe which
abounds with structure, galaxies, clusters of
galaxies, superclusters, voids and great walls (\cite{structure}).

\section{Inflation}

As successful as the big-bang cosmology is, it suffers from a dilemma
involving initial data.  Extrapolating back, one finds that
the Universe apparently began from a very special state:
A slightly inhomogeneous and very flat Robertson-Walker spacetime.
Collins and Hawking showed that the
set of initial data that evolve to a spacetime that is
as smooth and flat as ours is today is of measure zero (\cite{collins}).
[In the context of simple grand unified theories, the hot big-bang model
suffers from another serious problem:  the extreme overproduction of superheavy
magnetic monopoles (\cite{preskill}); in fact, attempting to solve the
monopole problem led Guth to inflation.]

The cosmological appeal of inflation is the lessening of the dependence
of the present state of the Universe upon its initial state.  Two
elements are essential:  (1) accelerated (``superluminal'')
expansion and the concomitant growth of the
scale factor; and (2) massive entropy production (\cite{htw}).
Through inflation, a small, smooth subhorizon-sized
patch of the early Universe grows to a large enough size and comes to contain
enough heat (entropy in excess of $10^{88}$) to encompass our
present Hubble volume.  In addition, superluminal expansion
guarantees that the Universe today appears flat (just
as any small portion of the surface of a large sphere appears flat).

While there is presently no standard model of inflation---just as
there is no standard model for physics at these energies
(typically $10^{15}\GeV$ or so)---viable models have much in
common.  They are based upon well posed, albeit highly speculative,
microphysics involving the classical evolution of a scalar field.
A nearly exponential expansion is driven by the potential
energy (``vacuum energy'') that arises when the scalar field
is displaced from its potential-energy minimum.  Provided
the potential is flat, during the time it takes for the field to roll
to the minimum of its potential the Universe undergoes many e-foldings
of expansion (more than around 60 or so are required to realize
the beneficial features of inflation).
As the scalar field nears the minimum,  the vacuum energy has been
converted to coherent oscillations of the scalar field, which
correspond to nonrelativistic scalar-field particles.  The eventual
decay of these particles into lighter particles and their thermalization
results in the ``reheating'' of the Universe and accounts for all the
heat in the Universe today (the entropy production event).

The growth of the cosmic scale
factor (by a factor greater than that since the end of inflation)
allows quantum fluctuations excited on very small scales ($\la 10^{-23}\rcm$)
to be stretched to astrophysical scales ($\ga 10^{25}\rcm$).
Quantum fluctuations in the scalar field responsible for inflation
ultimately lead to an almost scale-invariant spectrum
of density perturbations
(\cite{scalar1}; \cite{scalar2}; \cite{scalar3}; Bardin, Steinhardt \& Turner
1983),
and quantum fluctuations in
the metric itself lead to an almost scale-invariant
spectrum of gravity-waves
(\cite{tensor1}; Fabbri \& Pollock 1983; \cite{tensor3}; \cite{tensor4}).
Scale invariance for density perturbations means
scale-independent fluctuations in the gravitational potential
(equivalently, density perturbations of different wavelength
cross the horizon with the same amplitude);
scale invariance for gravity waves means that
gravity waves of all wavelengths cross the horizon with the same amplitude.
Because of subsequent evolution, neither the scalar nor the
tensor perturbations are scale invariant today.

\subsection{Grander Implications}

Inflation lessens greatly the ``specialness'' problem, but does
not eliminate it (Turner \& Widrow 1986; \cite{nohair2}; \cite{nohair3}).
All open FRW models will inflate and become flat; however,
many closed FRW models will recollapse before they can inflate.
If one imagines the most general initial spacetime as being comprised
of negatively and positively curved FRW (or Bianchi) models that are
stitched together, the failure of the positively
curved regions to inflate is of little consequence:  because
of exponential expansion during inflation the negatively curved
regions will occupy most of the space today.
Inflation does not solve the smoothness problem forever;
it just postpones the problem into the exponentially distant future:
We will be able to see outside our smooth inflationary
patch at a time $t\sim t_0
\exp [3(N-N_{\rm min}]$, where $N$ is the actual number of e-foldings
of inflation and $N_{\rm min}\sim 60$ is the minimum required to
solve the horizon/flatness problems.

Linde has emphasized that inflation has changed our view
of the Universe in a very fundamental way (\cite{eternal}).
While cosmologists have long used
the Copernician principle to argue that the Universe must be smooth
because of the smoothness of our Hubble volume, in the post-inflation
view, our Hubble volume is smooth because it is a small
part of a region that underwent inflation.  On
the largest scales the structure of the Universe is likely to be
very rich:  Different regions may have undergone different amounts of
inflation, may have different realizations of the
laws of physics because they
evolved into different vacuum states (of equivalent energy), and
may even have different numbers of spatial dimensions.  Since it is
likely that most of the volume of the Universe is still undergoing
inflation and that inflationary patches are being constantly produced
(eternal inflation), the age of the Universe is
a meaningless concept and our expansion age merely measures the
time back to our big bang -- the nucleation of our inflationary bubble.

\subsection{Specifics}

Guth's seminal paper (\cite{guth}) both introduced the idea of inflation
and showed that the model that he based the idea
upon did not work!  Thanks to very important contributions
by Linde (\cite{linde}) and Albrecht and Steinhardt (\cite{as}) that was
quickly
remedied, and today there are many viable models of inflation.
That of course is good news and bad news -- since it means that there
is no standard model of inflation.  The absence of a standard
model of inflation should of course be viewed in the light of our general
ignorance about physics at unification-scale energies.

Many different approaches have taken in constructing particle-physics
models for inflation.  Some have focussed on very simple scalar potentials,
e.g., $V(\phi ) = \lambda \phi^4$ or $=m^2\phi^2/2$, without regard
to connecting the model to any underlying theory (\cite{chaotic};
\cite{pjsmst}).
Others have proposed more complicated models that attempt to make
contact with speculations about physics at very high energies,
e.g., grand unification (\cite{pi1}; \cite{pi2}) supersymmetry (Holman, Ramond
\& Ross
1984;
\cite{olive}; \cite{lbl}; \cite{lisa})
preonic physics (Cvetic et al. 1990), or supergravity (\cite{liddle}).
Several authors have attempted
to link inflation with superstring theory
(\cite{banks1}; Brustein \& Veneziano 1994; \cite{banks3}) or ``generic
predictions'' of superstring theory such as pseudo-Nambu-Goldstone
boson fields (\cite{unnatural}).  While the scale of the vacuum energy
that drives inflation is typically of order $(10^{15}\GeV)^4$, a
model of inflation at the electroweak scale, vacuum energy $\approx(1\TeV )^4$,
has been proposed (\cite{knox}).  There are also models in which there are
multiple epochs of inflation (\cite{multiple1}; \cite{multiple2}).

In all of the models above gravity is described by
general relativity.  A qualitatively different approach is to
consider inflation in the context of alternative theories of
gravity.\footnote{After all, inflation probably involves physics at
energy scales not too different from the Planck scale and the
effective theory of gravity at these energies could well be
very different from general relativity; in fact, there are some
indications from superstring theory that gravity in these
circumstances might be described by a Brans-Dicke-like theory.}
The most successful of these models is
first-order inflation (\cite{lapjs}; Kolb 1991).  First-order inflation returns
to Guth's original idea of a strongly first-order
phase transition; in the context of general relativity Guth's model
failed because the phase transition, if inflationary, never completed.
In theories where the effective strength of gravity evolves, like
Brans-Dicke theory, the weakening of gravity during inflation
allows the transition to complete.  In other models based upon
nonstandard gravitation theory, the scalar field responsible for
inflation is itself related to the size of additional spatial dimensions,
and inflation then also explains why our three spatial dimensions are
so big, while the other spatial dimensions are so small.

All models of inflation have one feature in common:  the scalar
field responsible for inflation has a very flat potential-energy
curve and is very weakly coupled.  Invariably, this leads to a very
small dimensionless number, usually a coupling constant
of the order of $10^{-14}$.
Such a small number, like other small numbers in physics (e.g.,
the ratio of the weak to Planck scales $\approx 10^{-17}$ or
the ratio of the mass of the electron to the $W/Z$ boson masses $\approx
10^{-5}$), runs counter to one's belief that a truly fundamental
theory should have no tiny parameters, and cries out for an
explanation.  At the very least, this small number must be stabilized against
quantum corrections---which it is in all of the previously
mentioned models.\footnote{It is sometimes stated that inflation
is unnatural because of the small coupling of the scalar field
responsible for inflation; while the small coupling certainly begs
explanation, these inflationary models are not unnatural in
the technical sense as the small number is stable
against quantum fluctuations.}  In some models, the small number in the
inflationary potential is related to other small numbers in
particle physics:  for example, the ratio of the electron mass
to the weak scale or the ratio of the unification scale to
the Planck scale.  Explaining the origin of
the small number that seems to be associated with
inflation is both a challenge and an opportunity.

Because of the growing base of observations that bear on inflation,
another approach to model building is emerging:  the use of observations
to constrain the underlying inflationary model.  In Section 4 I will
discuss the possibilities for reconstructing the inflationary potential.

\subsection{Three Robust Predictions}

While there are many
varieties of inflation, there are three robust predictions
which are crucial to sharply testing inflation.\footnote{Because theorists
are so clever, it is not possible nor prudent to use the word
immutable.  Models that violate any or all of these ``robust
predications'' can and have been constructed.}

\begin{enumerate}

\item {\bf Flat universe.}  Because solving the ``horizon''
problem (large-scale smoothness in spite of small particle
horizons at early times) and solving the ``flatness'' problem
(maintaining $\Omega$ very close to unity until the present epoch)
are linked geometrically (\cite{eu}; \cite{htw}), this is the most robust
prediction of inflation.  Said another way, it is the prediction
that most inflationists would be least willing to give up.
[Even so, models of inflation have been
constructed where the amount of inflation is tuned just to give
$\Omega_0$ less than one today (Bucher, Goldhaber \& Turok 1995; \cite{pu2}).]
Through the Friedmann
equation for the scale factor, flatness implies that the total
energy density (matter, radiation, vacuum energy, and anything else) is
equal to the critical density.

\item {\bf Nearly scale-invariant spectrum of gaussian density perturbations.}
Essentially all inflation models predict a nearly scale-invariant
spectrum of gaussian density perturbations.  Described in terms
of a power spectrum, $P(k) \equiv \langle |\delta_k|^2 \rangle
= Ak^n$, where $\delta_k$ is the Fourier transform of the primeval
density perturbations, and the spectral index $n = 1$
in the scale-invariant limit.  The overall amplitude $A$
is model dependent.  Density perturbations give rise to CBR anisotropy
as well as seeding structure formation.
Requiring that the density perturbations are consistent
with the observed level of anisotropy of the CBR (and large enough
to produce the observed structure formation) is the most severe
constraint on inflationary models and leads to the small
dimensionless number that all inflationary models have.

\item {\bf Nearly scale-invariant spectrum of gravitational waves.}
These gravitational waves have wavelengths from around $1\km$
to the size of the present Hubble radius and beyond.  Described in
terms of a power spectrum for the dimensionless gravity-wave amplitude
at early times, $P_T(k) \equiv \langle |h_k|^2 \rangle = A_Tk^{n_T-3}$, where
the spectral index $n_T = 0$ in the scale-invariant limit.
Once again, the overall amplitude $A_T$ is model dependent (varying
as the value of the inflationary vacuum energy).  Unlike density
perturbations, which are required to initiate structure formation,
there is no cosmological lower bound to the amplitude of
the gravity-wave perturbations.  Tensor perturbations
also give rise to CBR anisotropy; requiring that they do not lead to
excessive anisotropy implies that the energy density that drove
inflation must be less than about $(10^{16}\GeV )^4$.  This
indicates that if inflation took place, it did so at an energy well
below the Planck scale.\footnote{To be more precise, the part of inflation
that led to perturbations on scales within the present horizon involved
subPlanckian energy densities.  In
some models of inflation, the earliest stage of inflation, which only
influences scales much larger than the present horizon,
involve energies as large as the Planck energy density.}

\end{enumerate}

There are other interesting consequences of inflation that
are less generic.  For example, in first-order inflation,
where reheating occurs through the nucleation and collision of
vacuum bubbles, there is an additional, larger amplitude, but
narrow-band, spectrum of gravitational waves ($\Omega_{\rm GW}h^2
\sim 10^{-6}$) (\cite{vacuumpop1}; Kosowsky, Turner \& Watkins 1992).
In other models large-scale primeval magnetic
fields of interesting size are seeded during inflation (\cite{bfield1};
\cite{bfield2}).

I want to emphasize the importance
of the tensor perturbations.   The attractiveness of a flat Universe
with scale-invariant density perturbations was appreciated long before
inflation.  Verifying these two predictions of inflation, while
important, will not provide a ``smoking gun.''  A spectrum
of nearly scale-invariant tensor perturbations is a defining feature
of inflation, and further, is crucial
to obtaining information about the underlying scalar potential.

CBR anisotropy probably provides the best possibility of
detecting the tensor perturbations, but their contribution
to CBR anisotropy has to be separated that of the
scalar perturbations.  Because the sky is
finite, sampling variance sets a fundamental limit:  the tensor
contribution to CBR anisotropy can only be separated from that of
the scalar if $T/S$ is greater than about $0.14$ (\cite{knoxmst})
($T$ is the contribution of tensor perturbations to the variance
of the CBR quadrupole and $S$ is the same for scalar perturbations).

It is possible that the stochastic background of gravitational
waves itself can be directly detected, though it appears that
the LIGO facilities being built will lack the sensitivity and
even space-based interferometry (e.g., LISA) is not a sure bet
(Turner, Lidsey \& White 1993;
\cite{tlw2}).

\section{Cold Dark Matter}

Cold dark matter actually draws from three important ideas --
inflation, big-bang nucleosynthesis, and the quest to
better understand the fundamental forces and particles.
As discussed above, inflation predicts a flat Universe
(total energy density equal to the critical density) and
nearly scale-invariant density perturbations.
Big-bang nucleosynthesis provides the most precise
determination of the density of ordinary matter,
present density between $1.5\times 10^{-31} \gcmm3$ and $4.5
\times 10^{-31}\gcmm3$, or fraction of critical
density $\Omega_B = 0.008h^{-2} - 0.024h^{-2}$, where
$H_0 = 100h \kms \Mpc^{-1}$ (\cite{bbn}).  Allowing $h=0.4 -0.8$,
 consistent with modern measurements (Fukugita, Hogan \& Peebles 1993;
 \cite{h0rev2}),
implies that ordinary matter can contribute at most 10\% of
the critical density.  If the inflationary prediction
is correct, then most of the matter in the Universe must be nonbaryonic.

This idea has received indirect support from particle physics.
Attempts to further our understanding of the particles
and forces have led to the prediction
of new, stable or long-lived particles that interact very feebly with ordinary
matter.  These particles, if they exist, should have
been present in great numbers during the earliest moments
and remain today in numbers sufficient to contribute
the critical density (\cite{pdm}).  Two of the most attractive
possibilities behave like cold dark matter:  a neutralino of mass
$10\GeV$ to $1000\GeV$, predicted in supersymmetric theories, and an axion
of mass $10^{-6}\eV$ to $10^{-4}\eV$, needed to solve a subtle problem of the
standard model of particle physics (strong-CP problem).
The third interesting possibility
is that one of the three neutrino species has a mass between $5\eV$ and
$30\eV$; neutrinos move very fast and are referred to
as hot dark matter.\footnote{The possibility that most of the exotic
particles are fast-moving neutrinos -- hot dark matter -- was explored first
and found to be inconsistent with observations (\cite{hdm}).  The
problem is that large structures form first and must fragment into
smaller structures, which conflicts with the fact that
large structures are just forming today.}

According to cold dark matter theory CDM particles
provide the cosmic infrastructure:  It is their gravitational
attraction that forms and holds cosmic structures together.
Structure forms in a hierarchical manner, with galaxies forming
first and successively larger objects forming thereafter (\cite{faberetal}).
Quasars and other rare objects form at redshifts of up to
five, with ordinary galaxies forming a short time later.  Today, superclusters,
objects made of several
clusters of galaxies, are just becoming bound by the gravity of
their CDM constituents.  The formation of larger and larger objects
continues.  In the clustering process regions of space are left
devoid of matter -- and galaxies -- leading to voids.

\subsection{Standard Cold Dark Matter}

When the cold dark matter scenario emerged more than
a decade ago many referred to it as a no parameter
theory because it was so specific compared to previous
models for the formation of structure.  This was an overstatement
as there are cosmological quantities that must be
known to determine the development of structure in detail.
However, the data available did not require precise knowledge of
these quantities to test the model.

Broadly speaking these parameters can be organized
into two groups.  First are the cosmological parameters:  the
Hubble constant, specified by $h$; the density of
ordinary matter, specified by $\Omega_B h^2$; the power-law index $n$ and
normalization $A$ that quantify the density
perturbations; and the amplitude and spectral index
$n_T$ that quantify the gravitational waves.
The inflationary parameters fall into this category
because there is no standard model of inflation.
On the other hand, once determined they can be used to discriminate
between models of inflation.

The other quantities specify the composition of invisible matter
in the Universe:  radiation, dark matter, and a possible cosmological
constant.  Radiation refers to relativistic
particles:  the photons in the CBR, three massless
neutrino species (assuming none of the neutrino species has
a mass), and possibly other undetected relativistic particles
(some particle-physics theories predict the existence of additional
massless particle species).  At present relativistic particles
contribute almost nothing to the energy density in the Universe,
$\Omega_R \simeq 4.2 \times 10^{-5}h^{-2}$; early on --- when
the Universe was smaller than about $10^{-5}$ of its present
size --- they dominated the energy content.

In addition to CDM particles, the dark matter could include other particle
relics.  For example, each neutrino species has a number density of
$113\cmm3$, and a neutrino species of mass $5\eV$
would account for about 20\% of the critical density
($\Omega_\nu = m_\nu/90h^{2}\eV$).  Predictions for
neutrino masses range from $10^{-12}\eV$ to several MeV, and
there is some experimental evidence that at least one of
the neutrino species has a small mass
(\cite{numass1}; \cite{numass2}; \cite{numass3};
\cite{numass4};
\cite{numass5}; \cite{numass6}).

Finally, there is the cosmological constant.  Both introduced
and abandoned by Einstein, it is still with us.
In the modern context it corresponds
to an energy density associated with the quantum vacuum.  At present,
there is no reliable calculation of the value that the cosmological
constant should take, and so its existence
must be regarded as a logical possibility (\cite{cosmoconst}).

The original no parameter cold dark matter model,
referred to as standard CDM, is characterized by:  $h=0.5$, $\Omega_B =0.05$,
$\Omega_{\rm CDM}=0.95$, $n=1$, no gravitational waves and
standard radiation content.  The overall normalization of the
density perturbations was fixed by
comparing the predicted level of inhomogeneity with that
seen today in the distribution of bright galaxies.
Specifically, the amplitude $A$ was
determined by comparing the expected mass fluctuations in
spheres of radius $8h^{-1}\Mpc$ (denoted by $\sigma_8$) to the
galaxy-number fluctuations in spheres of the same size.
The galaxy-number fluctuations on the scale $8h^{-1}\Mpc$
are unity; adjusting $A$ to achieve $\sigma_8 = 1$ corresponds
to the assumption that light, in the form of bright galaxies, traces mass.
Choosing $\sigma_8$ to be less than one means that light is more
clustered than mass and is a biased tracer of mass.
There is some evidence that bright galaxies are somewhat
more clumped than mass with biasing factor $b\equiv 1/\sigma_8
\simeq 1 - 2$ (\cite{biasing}).

\begin{figure}
\center
\leavevmode
\epsfig{figure=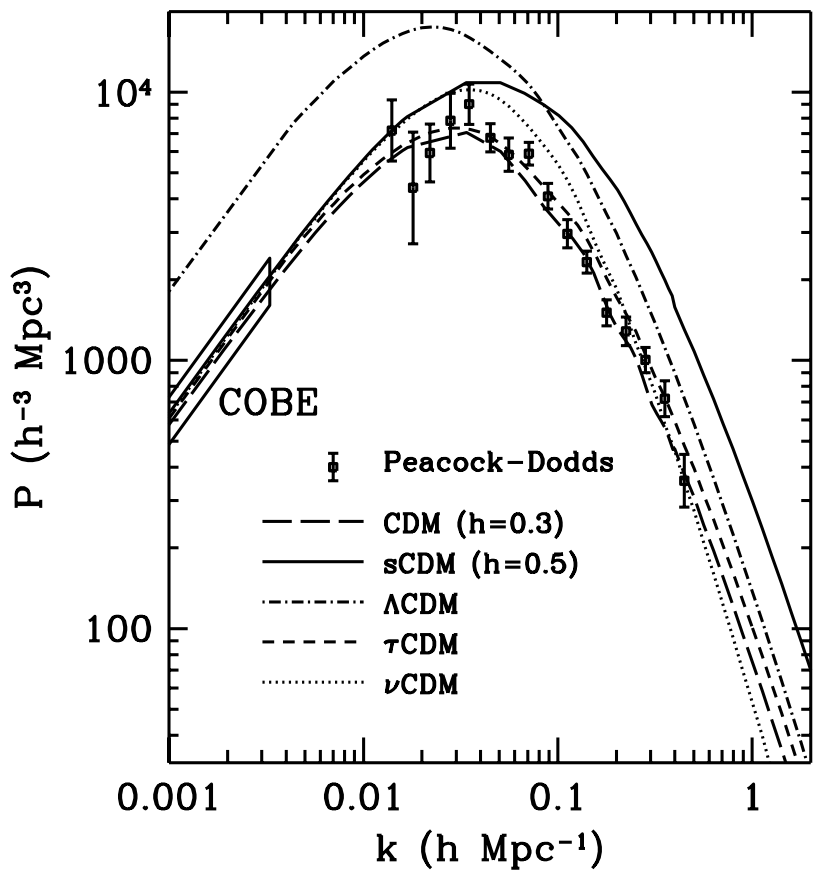,height=4.5in}
\caption{Measurements of the power spectrum,
$P(k) \equiv |\delta_k|^2$, and the predictions of different
COBE-normalized CDM models.
The points are from several redshift surveys as analyzed
in Ref. (~\protect\cite{pd}); the models are:  $\Lambda$CDM
with $\Omega_\Lambda =0.6$ and $h=0.65$; standard CDM (sCDM),
CDM with $h=0.35$; $\tau$CDM (with the energy equivalent
of 12 massless neutrino species) and $\nu$CDM with $\Omega_\nu = 0.2$
(unspecified parameters have their standard CDM values).}
\end{figure}

A dramatic change occurred with the detection of CBR anisotropy
by COBE in 1992 (\cite{dmr}).
The COBE measurement permitted a precise normalization of the
amplitude of density perturbations on very large scales
($\lambda \sim 10^4h^{-1}\Mpc$) without regard to the issue of biasing.
[CBR anisotropy on the angular scale $\theta$ arises primarily due to
inhomogeneity on length scales $\lambda \sim 100h^{-1}
\Mpc (\theta /{\rm deg})$.]
For standard CDM, the COBE normalization leads to:  $\sigma_8
=1.2 \pm 0.1 $ or anti-bias since $b=1/\sigma_8\simeq 0.7$.  The
pre-COBE normalization ($\sigma_8 = 0.5$) led to too little power
on scales of $30h^{-1}\Mpc$ to $300h^{-1}\Mpc$, as compared
to what was indicated in redshift surveys, the angular correlations
of galaxies on the sky and the peculiar velocities of galaxies.
The COBE normalization leads to about the right amount of power on
these scales, but predicts too much power on small
scales ($\la 8 h^{-1}\Mpc$); see Fig.~1.

While standard CDM is in general agreement with the observations, a
consensus has developed that the conflict just mentioned is probably
significant (\cite{jpo-ll1}; \cite{jpo-ll2}).  This has led to a new look at
the cosmological and invisible-matter parameters and
to the realization that the problems of standard CDM
are simply a poor choice for the standard parameters.

\section{A Flood of Data}

Standard CDM has served well to focus everyone's attention on the same problem.
However, the quality and quantity of data have improved
and knowledge of the cosmological and invisible-matter parameters
has become important for serious testing of CDM and inflation.
There are a variety of
combinations of the parameters that
lead to good agreement with the existing data on both large
and small length scales -- and thus can make a claim to being
the new standard CDM model.  Figure 2 shows the allowed values of
the cosmological parameters for several COBE-normalized CDM
models.\footnote{Computation
of both the CBR anisotropy and the level of inhomogeneity
today depends upon the invisible-matter content and the cosmological
parameters and requires that the distribution of
matter and radiation be evolved numerically; for
details see
Refs.~(\cite{boltzmann1}; \cite{boltzmann2}; \cite{boltzmann3};
\cite{boltzmann4}; \cite{boltzmann5}; \cite{boltzmann6}; \cite{boltzmann7}).
The discussion of viable models is
a summary of collaborative work (\cite{dgt}).}

More precisely, for a given CDM model --- specified by
the cosmological and invisible-matter parameters ---
the expected CBR anisotropy is computed and required to
be consistent with the four-year COBE data set at
the two-sigma level (\cite{dmr4yr1}; \cite{dmr4yr2}; \cite{dmr4yr3}).
The expected level
of inhomogeneity in the Universe today is compared to three robust
measurements of inhomogeneity:  the shape of the power spectrum as
inferred from surveys of the distribution of galaxies today (\cite{pd});
a determination of $\sigma_8$ based upon the
abundance of rich, x-ray emitting clusters (\cite{sigma8cluster});
and the abundance of hydrogen clouds at high redshift
(which probes early structure formation) (Lanzetta, Wolfe \& Turnshek 1995;
\cite{Dlya2}).

\begin{figure}
\center
\leavevmode
\epsfig{figure=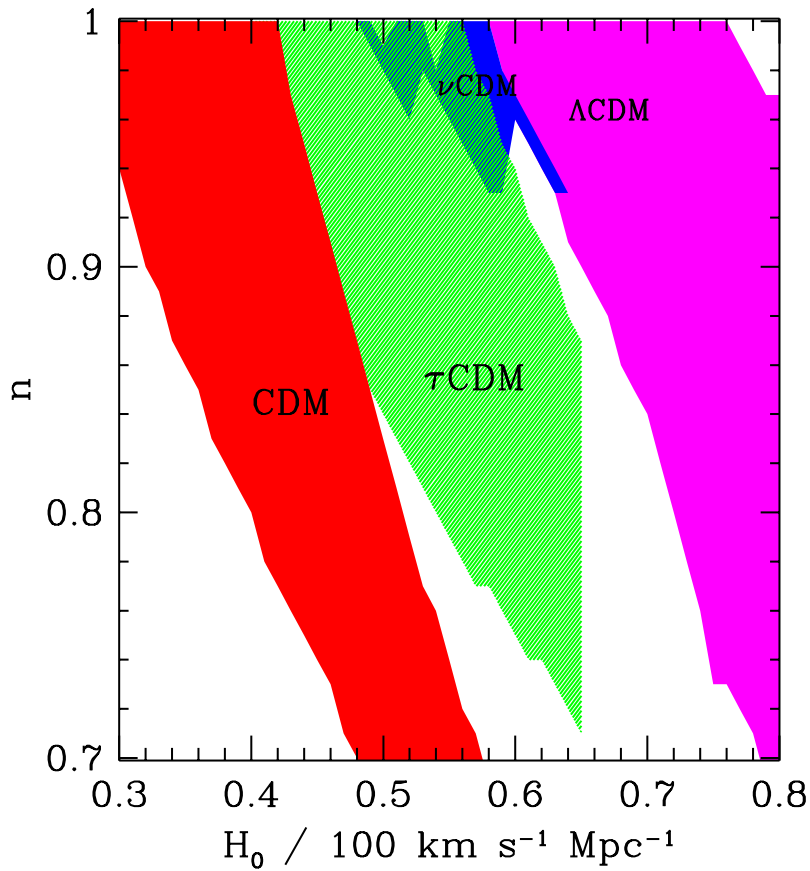,height=4.5in}
\caption{Acceptable values of the cosmological
parameters $n$ and $h$ for CDM models with standard
invisible-matter content (CDM), with 20\% hot dark matter ($\nu$CDM),
with additional relativistic particles (the energy equivalent
of 12 massless neutrino species, denoted $\tau$CDM), and with a cosmological
constant that accounts for 60\% of the critical density ($\Lambda$CDM).
Note that standard CDM ($n=1$ and $h=0.5$) is not viable.}
\end{figure}

Figure 2 summarizes the overall picture.  The simplest CDM models --
those with standard invisible-matter content -- lie in a
region that runs diagonally from smaller Hubble constant
and larger $n$ to larger Hubble constant and smaller $n$.  That is, higher
values of the Hubble constant require more tilt (tilt referring
to deviation from scale invariance).
Note too that standard CDM is well outside of the allowed range.
Current measurements of CBR anisotropy on the degree scale,
as well as the COBE four-year anisotropy data,
preclude $n$ less than about 0.7 (see Fig.~3).  This implies that the
largest Hubble constant consistent with the simplest CDM
models is slightly less than $60\kms\Mpc^{-1}$.
If the invisible-matter content is nonstandard, higher values of the
Hubble constant can be accommodated.  In Fig.~2, $\Omega_\nu$ is
taken to be 0.2; in fact, this is essentially the largest value
allowed by measurements of the power spectrum (\cite{dgt}; \cite{nucontent1};
\cite{nucontent2}).
On the other hand, even $\Omega_\nu = 0.05$ (around $1\eV$ worth
of neutrinos) can have important consequences (e.g., accommodating
a higher value of the Hubble constant or more nearly scale-invariant
density perturbations).

\begin{figure}
\center
\leavevmode
\epsfig{figure=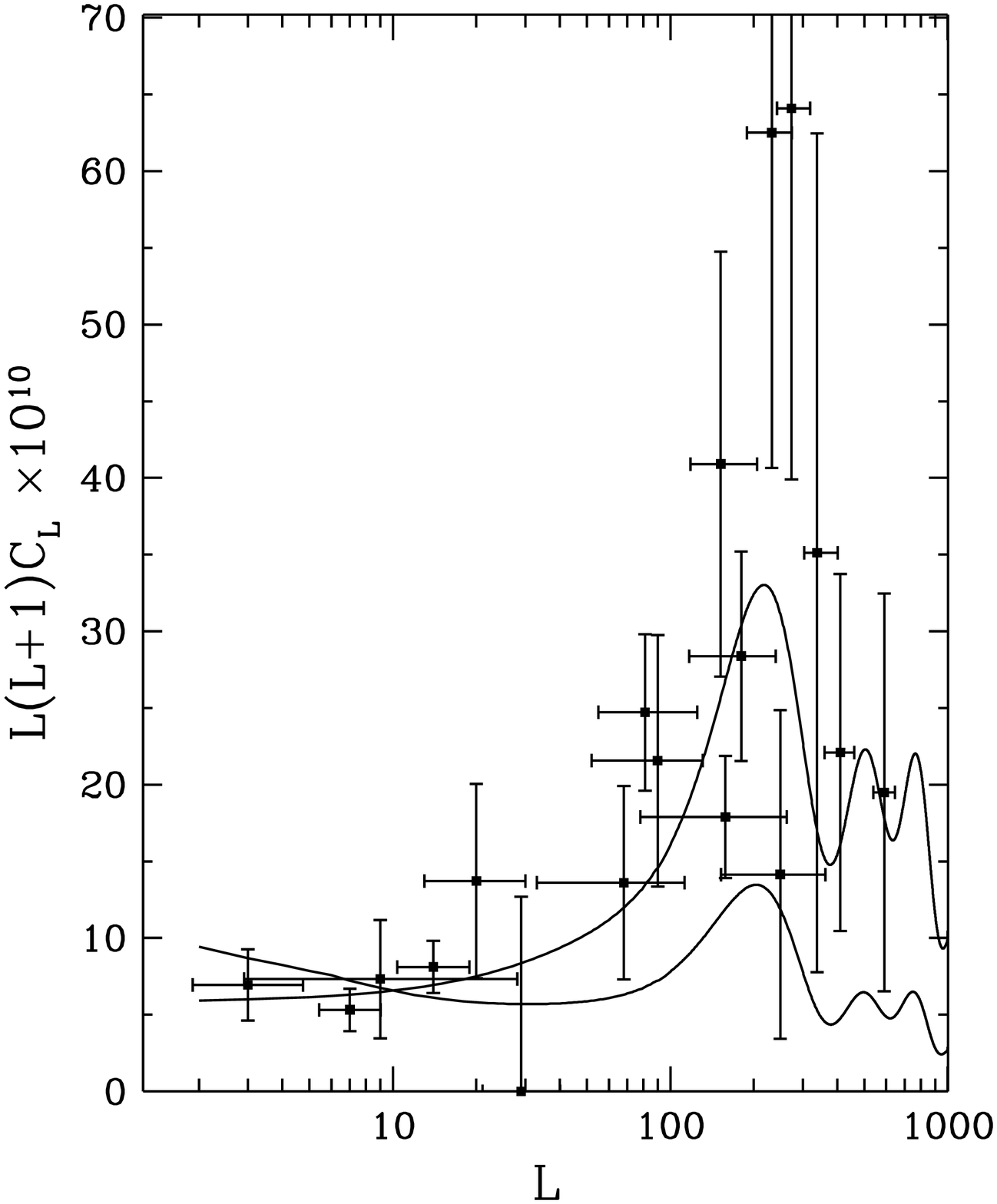,height=4.5in}
\caption{Summary of CBR anisotropy measurements
and predictions for three CDM models [adapted from
Ref.~(\protect\cite{cbrsummary})].
Plotted are the squares of the measured multipole amplitudes ($C_l = \langle
|a_{lm}|^2\rangle$) in units of $7\times 10^{-10}$ vs. multipole number $l$.
The temperature difference on angular scale $\theta$ is
given roughly by $\protect\sqrt{l(l+1)C_l}$ with $l\sim 200^\circ /\theta$.
The theoretical curves are standard CDM and CDM with $n=0.7$ and $h=0.5$.}
\end{figure}

Changes in the different parameters from their standard CDM values
alleviate the excess power on small scales in different ways.
Tilt has the effect of reducing
power on small scales when power on very large scales is fixed by COBE.
A small admixture of hot dark matter works because
fast moving neutrinos suppress the growth
of inhomogeneity on small scales by streaming from
regions of higher density and to regions of lower density.
(It was in fact this feature of hot dark matter that led to
the demise of the hot dark matter model for structure formation.)

A low value of the Hubble constant, additional radiation
or a cosmological constant all reduce power on small scales by
lowering the ratio of matter to radiation.  Since the
critical density depends upon the square of the
Hubble constant, $\rho_{\rm Critical} = 3H_0^2/8\pi G$,
a smaller value corresponds to a lower matter density
since $\rho_{\rm Matter} = \rho_{\rm Critical}$ for a flat
Universe without a cosmological constant.
Shifting some of the critical density to vacuum energy
also reduces the matter density since $\Omega_{\rm Matter} = 1 -
\Omega_\Lambda$.  Lowering the ratio of matter to radiation
reduces the power on small scales in a subtle way.  While the primeval
fluctuations in the gravitational potential are nearly scale-invariant,
density perturbations today are not because the Universe
made a transition from an early radiation-dominated phase ($t\la
1000\,$yrs), where the growth of density perturbations is inhibited,
to the matter-dominated phase, where growth proceeds unimpeded.
This introduces a feature in the power spectrum today
(see Fig.~1), whose location depends upon the relative amounts
of matter and radiation.  Lowering the ratio of matter
to radiation shifts the feature to larger scales and with power
on large scales fixed by COBE this leads to less power on small scales.

Some of the viable models have been discussed
as singular solutions
--- cosmological constant (\cite{lcdm1}; \cite{pdm}; \cite{lcdm3}; Efstathiou
et
al. 1990; \cite{lcdm5}),
very low Hubble constant (\cite{lhccdm}),
tilt (\cite{tcdm1}; \cite{tcdm2}; \cite{tcdm3}; \cite{tcdm4};
\cite{tcdm5}; \cite{tcdm6}; \cite{tcdm7}; \cite{tcdm8}),
tilt + low Hubble constant (\cite{stp}),
extra radiation (Dodelson, Gyuk \& Turner 1994; \cite{taucdm2}),
an admixture of hot dark matter (Shafi \& Stecker 1984;
\cite{nucdm2}; \cite{nucdm3}; Pogosyan \& Starobinsky 1995).  There
is actually a continuum of viable models, as can be seen in Fig.~2,
which arises because of imprecise knowledge of cosmological
parameters and the invisible-matter sector and not the
inventiveness of theorists.

\subsection{Other and Future Considerations}

There are many other observations that bear on structure
formation.  However, with cosmological data
systematic error and interpretational
issues are important considerations.  In fact,
if all extant observations
were taken at face value, there is no viable model for
structure formation, cold dark matter or otherwise!
With this as a preface, I now discuss some of the
other existing data as well as future measurements that
will more sharply test cold dark matter.

There is tension between measures of the age of the
Universe and determinations of the Hubble constant.
It arises because determinations of the ages of the oldest stars lie
between $13\Gyr$ and $19\Gyr$ (\cite{age1}; \cite{age2})
and recent measurements of the Hubble
constant favor values between $60\kms\Mpc^{-1}$ and
$80\kms\Mpc^{-1}$ (\cite{h_01}; \cite{h_02};
Freedman et al. 1994), which, for $\Omega_{\rm Matter} =1$,
leads to a time back to the bang of $11\Gyr$ or less (see Fig.~4).\footnote{The
time back to the bang depends upon $H_0$, $\Omega_{\rm Matter}$
and $\Omega_\Lambda$; for $\Omega_{\rm Matter}=1$ and $\Omega_\Lambda
=0$, $t_{\rm BB}
= {2\over 3}H_0^{-1}$, or $13\Gyr$ for $h=0.5$ and $10\Gyr$
for $h=0.65$.  For a flat Universe with a cosmological constant
the numerical factor is larger than 2/3 (see Fig.~4).}
These age determinations receive additional support from
estimates of the age of the galaxy based upon the decay of
long-lived radioactive isotopes and the cooling of white-dwarf
stars, and all methods taken together make a strong case for
an absolute minimum age of $10\Gyr$ (\cite{agereview}).  It should be noted
that within the uncertainties there is no inconsistency,
even for $\Omega_{\rm Matter} =1$.

\begin{figure}
\center
\leavevmode
\epsfig{figure=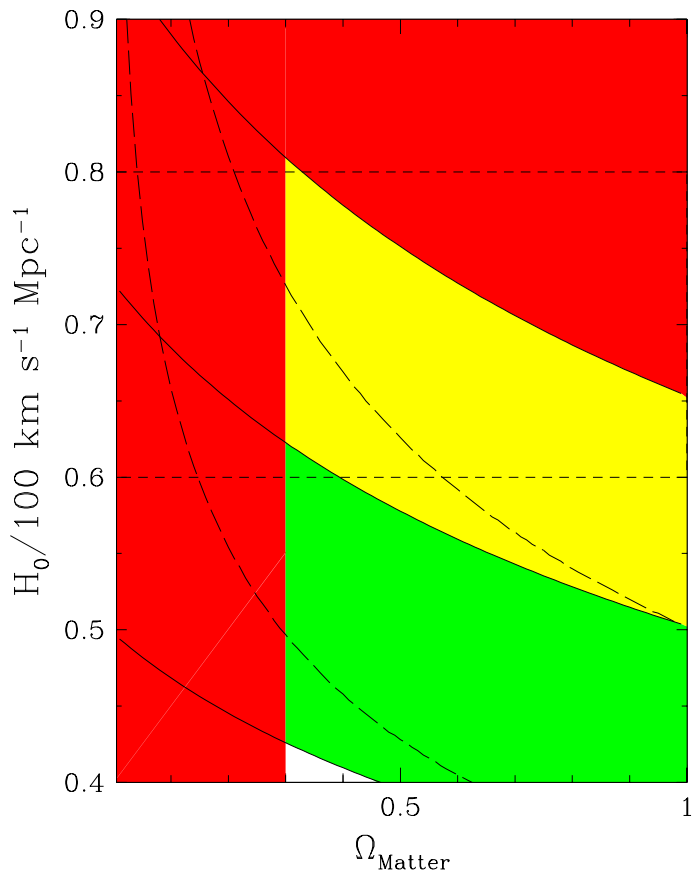,height=4.5in}
\caption{Isochrones in the $H_0$ - $\Omega_{\rm Matter}$
plane.  The bottom band corresponds to time back to the bang of
between $13\Gyr$ and $19\Gyr$; the lightest band to
between $10\Gyr$ and $13\Gyr$;  the darkest region is
disallowed:  $\Omega_{\rm Matter} < 0.3$ or expansion time less than
$10\Gyr$.  Broken horizontal lines indicate the range
favored for the Hubble constant, $80\kms\Mpc^{-1} > H_0 >
60\kms\Mpc^{-1}$.  The age -- Hubble
constant tension is clear, especially for the inflationary prediction
of $\Omega_{\rm Matter} = 1$.  The broken curves denote the
$13\Gyr - 19\Gyr$ isochrone for $\Lambda$CDM; a cosmological constant
greatly lessens the tension.}
\end{figure}

While age is not a major issue for cold dark matter --- large-scale
structure favors an older Universe by virtue of a lower Hubble
constant or cosmological constant (see Fig.~5)
--- the Hubble constant still
has great leverage.  If it is determined to be greater than about
$60\kms\Mpc^{-1}$, then only CDM models with nonstandard
invisible-matter content -- a cosmological constant or additional radiation
-- can be consistent with large-scale structure.  If
$H_0$ is greater than $65\kms\Mpc^{-1}$, consideration
of the age of the Universe
leaves $\Lambda$CDM as the lone possibility.  The issue
of $H_0$ is not settled, but the use of
Type Ia supernovae as standard candles, the study of
Cepheid variable stars in nearby galaxies using the Hubble
Space Telescope, and other methods make it likely that it will be soon.

\begin{figure}
\center
\leavevmode
\epsfig{figure=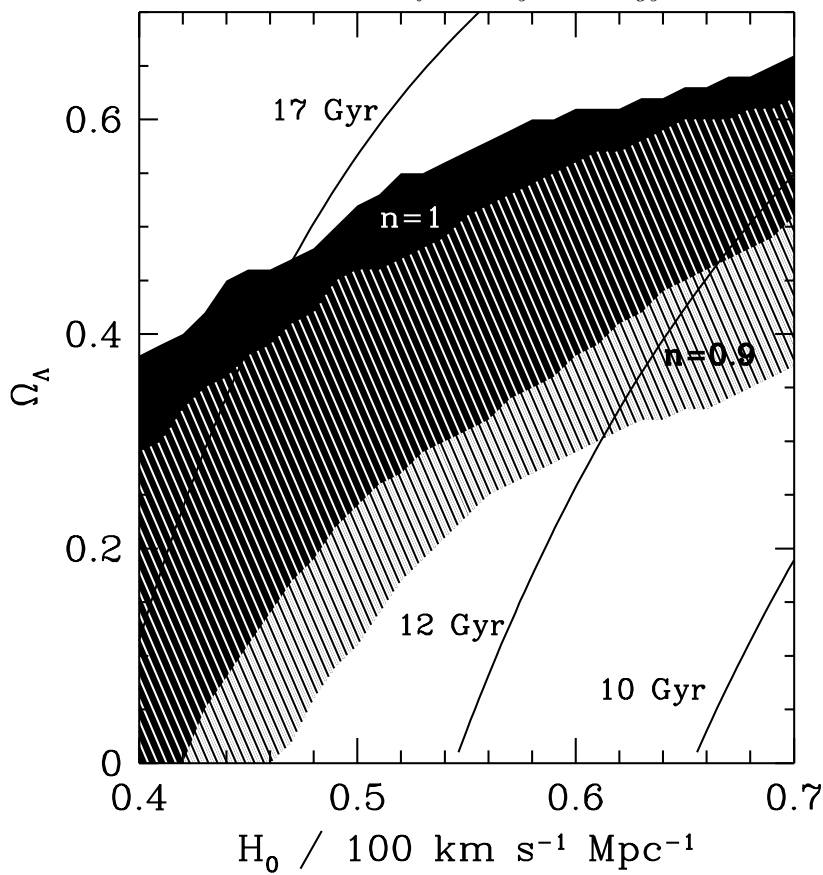,height=4.5in}
\caption{Acceptable values of $\Omega_\Lambda$
and $h$ for $n=0.9, 1.0$.  Note that large-scale structure
considerations generally favor a more aged Universe -- smaller
$h$ or larger $\Omega_\Lambda$.}
\end{figure}

If CDM is correct, baryons make up a small fraction of matter
in the Universe.  Most of the baryons in galaxy clusters are in the hot, x-ray
emitting intracluster gas and not the luminous galaxies.
The measured x-ray flux fixes the mass in baryons, while
the measured x-ray temperature fixes
the total mass (through the virial theorem).
The baryon-to-total-mass has been determined from x-ray measurements
for more than ten clusters and is found to be
$M_{\rm B}/M_{\rm TOT} \simeq (0.04-0.1)
h^{-3/2}$ (\cite{gasratio1}; \cite{gasratio2};
\cite{gasratio3}).  Because of their size,
clusters should represent a fair sample
of the cosmos and thus the baryon-to-total mass ratio should
reflect its universal value, $\Omega_B/\Omega_{\rm Matter}
\simeq (0.01 - 0.02)h^{-2}/\Omega_{\rm Matter}$.  These
two ratios are consistent for models with a very low
Hubble constant, $h\sim 0.4$ and $\Omega_{\rm Matter} = 1$,
or with a cosmological constant and $\Omega_{\rm Matter}\sim 0.3$.
However, important assumptions are made in this analysis
-- that the hot gas is unclumped and in virial
equilibrium and that magnetic fields do not provide significant
pressure support for the gas -- if any one of them is not valid the actual
baryon fraction would be smaller (\cite{outs}), allowing
for consistency with a larger value of $H_0$ without recourse
to a cosmological constant.

The halos of individual spiral galaxies like our own are not
large enough to provide a fair sample of matter in the
Universe -- for example, much
of the baryonic matter has undergone dissipation and condensed
into the disk of the galaxy.  Nonetheless, the content
of halos is expected to be primarily CDM particles.  This is
consistent with the fact that visible stars, hot gas, dust,
and even dark stars acting as microlenses (known as
MACHOs) account for only a fraction of the mass of our own
halo (\cite{macho1}; \cite{macho2}; \cite{nostars1}; \cite{nostars2}).

Determining the mean mass density of the Universe
would discriminate between models with and without a cosmological constant,
as well as test the inflationary prediction of a flat
Universe.  A definitive determination is still lacking.
The measurement that averages over the largest volume --
and thus is potentially most useful -- uses the
peculiar velocities of galaxies.  Peculiar velocities arise due to the
inhomogeneous distribution of matter, and the mean matter
density can be determined by relating the peculiar
velocities to the observed distribution of galaxies.  The results of this
technique
indicate that $\Omega_{\rm Matter}$ is at least 0.3 and perhaps as
large as unity (\cite{strauss}; \cite{dekel}).   Though not definitive,
this provides strong evidence for the existence of nonbaryonic dark
matter, a key aspect of cold dark matter.

A different approach to the mean density is through the deceleration
parameter $q_0$, which quantifies the slowing of the expansion
due to the gravitational attraction of matter in the Universe.
Its value is given by $q_0
 = {1\over 2}\Omega_{\rm Matter} - \Omega_\Lambda$ (vacuum energy actually
leads to accelerated expansion) and can be determined by relating the
distances and redshifts of distant objects.  In all but the $\Lambda$CDM
scenario, $q_0 =0.5$; for $\Lambda$CDM, $q_0 \sim -0.5$.
Two groups are trying to measure $q_0$ by using high redshift
($z\sim 0.4 - 0.7$) Type Ia supernovae as standard candles;
the preliminary results of one group suggest
that $q_0$ is positive (\cite{lblq_0}).
More than a dozen distant Type Ia supernovae were discovered
this year and both groups should soon have
enough to measure $q_0$ with a precision of $\pm 0.2$.

Gravitational lensing of distant QSOs by intervening galaxies
is another way to measure $q_0$, and the frequency of
QSO lensing suggests that $q_0 > -0.6$ (\cite{qsolensing}).
The distance to a QSO of given redshift is larger
for smaller $q_0$, and thus the probability for its being
lensed by an intervening galaxy is greater.

The 10\,m Keck Telescope and the Hubble Space Telescope are
providing the deepest images of the Universe ever
and are revealing
details of galaxy formation as well as the formation and
evolution of clusters of galaxies (\cite{fhp}).  The Keck has made the
first detection of deuterium in high redshift hydrogen
clouds (\cite{D1}; \cite{D2}; \cite{D3};
Tytler, Fan \& Burles 1996; \cite{D5}; \cite{D6};
\cite{D7}; Rugers \& Hogan 1996b; \cite{D9}).  This is a new confirmation of
big-bang nucleosynthesis and has the potential of pinning
down the density of ordinary matter to a precision of 10\%.

The level of inhomogeneity in the Universe today is determined largely
from redshift surveys, the largest of which contain of order $10^4$ galaxies.
Two larger surveys are in progress.  The Sloan Digital Sky Survey
will obtain positions for two-hundred million galaxies and redshifts
for one million over a quarter of the sky; the Anglo - Australian
Two-Degree Field will gather 250,000 redshifts in hundreds of
two-degree fields.  They
will allow the power spectrum to be measured more precisely
and out to large enough scales ($500h^{-1}\Mpc$)
to connect with measurements from CBR anisotropy on angular scales
of up to five degrees.

The most fundamental element of cold dark matter --
the existence of the CDM particles themselves -- is being tested.
While the interaction of CDM particles with ordinary matter occurs
through very feeble forces and makes their existence difficult to test,
experiments with sufficient sensitivity to
detect the CDM particles that hold our own galaxy together if they are
in the form of axions of mass $10^{-6}\eV - 10^{-4}\eV$ (\cite{llnl}) or
neutralinos of mass tens of GeV (\cite{neutralinos}) are now underway.
Evidence for the existence of the neutralino could also come
from particle accelerators searching for other supersymmetric particles.
In addition, several experiments sensitive to neutrino masses
are operating or are planned, ranging
from accelerator-based neutrino oscillation experiments
to the detection of solar neutrinos to the study of the
tau neutrino at $e^+e^-$ colliders.

CBR anisotropy probes the power spectrum most cleanly as
it is related directly to the distribution of matter
when density perturbations were very small (\cite{cbranisotropy}).
Current measurements are beginning to test CDM and
differentiate between the variants (see Fig.~3); e.g., a spectral
index $n<0.7$ is strongly disfavored.  More than ten groups
are making measurements with instruments in space, on balloons and
at the South Pole.  Two new satellite-borne experiments have been
approved:  NASA will launch MAP in 2001 and ESA will launch
COBRAS/SAMBA in 2004.  They will map the anisotropy of the CBR
with better than $0.2^\circ$ resolution (more than
30 times better than COBE).  The results from these maps should easily
discriminate between the different variants of CDM (see Fig.~6).

\begin{figure}
\center
\leavevmode
\epsfig{figure=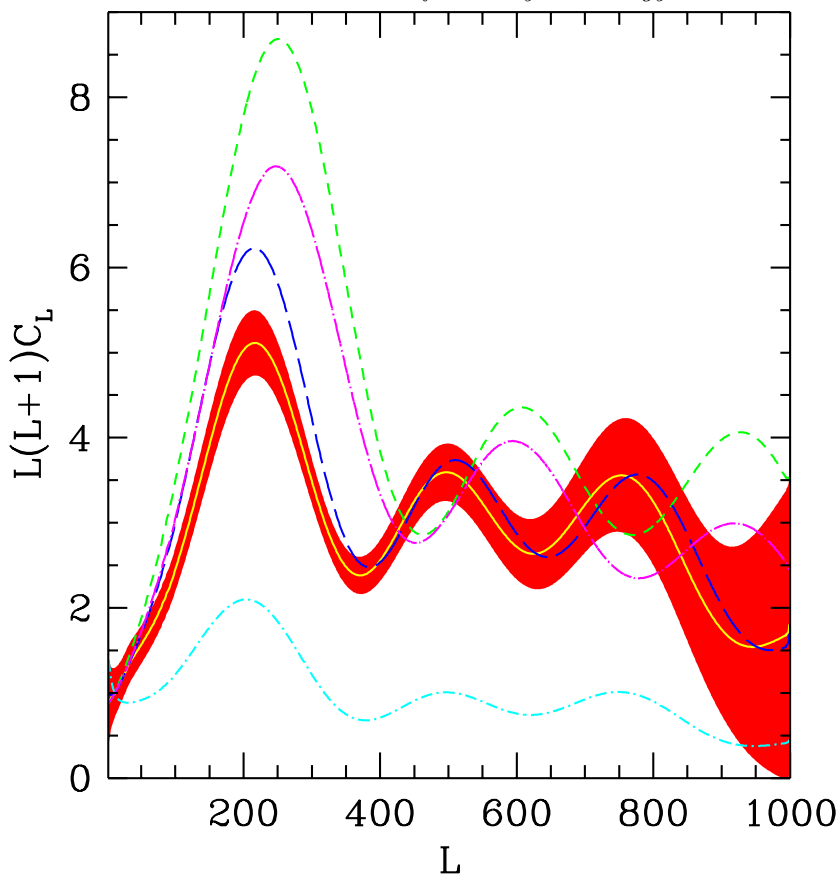,height=4.5in}
\caption{Predicted angular power spectra of
CBR anisotropy for several viable CDM models and the anticipated
uncertainty from a CBR satellite experiment with
angular resolution of $0.3^\circ$.  From top to bottom the models
are:  CDM with $h=0.35$, $\tau$CDM with the energy equivalent
of 12 massless neutrino species, $\Lambda$CDM with $h=0.65$ and
$\Omega_\Lambda = 0.6$, $\nu$CDM with $\Omega_\nu = 0.2$,
and CDM with $n=0.7$ (unspecified parameters have their standard CDM values).}
\end{figure}

The first and most powerful test to emerge from these measurements
will be the location of the first (Doppler) peak in the angular power
spectrum (see Fig. 6) (Kamionkowski, Spergel \& Sugiyama 1994;
\cite{omegacmb2}; \cite{omegacmb3}).  All
variants of CDM predict the location of the first peak to lie
in roughly the same place. On the other hand, in an open Universe
(total energy density less than critical)
the first peak occurs at a larger value of $l$ (much smaller
angular scale).  This will provide an important test of inflation.
In addition, theoretical studies (\cite{learn1}; \cite{learn2})
indicate that $n$ could be
determined to a precision of a few percent, $\Omega_\Lambda$
to ten percent, and perhaps even $\Omega_\nu$ to enough
precision to test $\nu$CDM (\cite{US}).

If {\it all} the current observations -- from recent Hubble
constant determinations to the cluster baryon fraction -- are taken at
face value, the cosmological constant + cold dark matter model
is probably the best fit (\cite{bestfit1};
\cite{bestfit2}), though there may be a conflict
with the measurement of $q_0$ with Type Ia supernovae.
$\Lambda$CDM raises a fundamental question -- the
origin of the implied vacuum energy, about $(10^{-2}\eV)^4$ --
since there is no known principle or mechanism that explains why
it is less than $(300\GeV )^4$, let alone $(10^{-2}\eV )^4$
(\cite{cosmoconst}).
In any case, it would be imprudent to
take all the observational data at face value because of important
systematic and interpretational uncertainties.  To paraphrase the
biologist Francis Crick, a theory that fits all the data at
any given time is probably wrong as some of the data are
probably not correct.

\subsection{Reconstruction}

If inflation and the cold dark matter theory are shown to be correct,
a window to the very early Universe ($t\sim 10^{-32}\sec$) will
have been opened.  While it is certainly premature to jump to this
conclusion, I would like to illustrate one example of what one could
hope to learn.  The spectra and amplitudes
of the the tensor and scalar metric perturbations predicted by
inflation depend upon the underlying model, to be specific, the
shape of the inflationary scalar-field potential.
If one can measure the power-law
index of the scalar spectrum and the amplitudes of the scalar
and tensor spectra, one can recover the value of the potential
and its first two derivatives around the point on the potential
where inflation took place (\cite{reconstruct1,reconstruct2}):
\begin{eqnarray}
V & = & 1.65 T\, {m_{\rm Pl}}^4  , \\
V^\prime & = & \pm \sqrt{8\pi r \over 7}\, V/{m_{\rm Pl}} , \\
V^{\prime\prime} & = & 4\pi \left[ (n-1) + {3\over 7} r \right]\,
V /{m_{\rm Pl}}^2 ,
\end{eqnarray}
where $r\equiv T/S$ ($T$ is the contribution of tensor
perturbations to the variance of the CBR quadrupole and
$S$ is the same for scalar perturbations), prime indicates derivative
with respect to $\phi$, $\mpl = 1.22\times 10^{19}\GeV$ is
the Planck energy, and the sign of $V^\prime$ is
indeterminate.  In addition, if the tensor spectral index $n_T$
can be measured a consistency relation, $n_T = -r/7$,
can be used to test inflation.
Reconstruction of the inflationary scalar potential would
shed light on the underlying physics of
inflation as well as physics at energies of the
order of $10^{15}\GeV$.

\section{Concluding Remarks}

Thanks to the work of Igor Novikov and others, the fifteen years following
the discovery of the CBR saw the establishment of a physical
basis for the expanding Universe, the hot big-bang cosmology.
With this as a firm foundation, the decade of the 1980s has
produced many bold and interesting
speculations about the earliest history of the Universe
which address the most fundamental
problems in cosmology.  Two of these ideas, inflation and cold dark matter,
are so attractive that experimenters and observers took them seriously.
The decade of the 1990s is producing a
flood of data that are testing inflation and cold dark matter.
The stakes for both cosmology and fundamental physics
are high:  if correct, inflation and cold dark matter
represent a major extension of the big bang and our understanding
of the Universe, which would certainly shed
light on fundamental physics at energies beyond the reach
of terrestrial accelerators.


\begin{thebibliography}{000}

\bibitem[Abbott \& Wise 1984]{tensor4}{\sc Abbott, L. \& Wise, M.} 1984
    {\it Nucl.\ Phys.\ B} {\bf 244}, 541.

\bibitem[Adams et al. 1993]{tcdm8}{\sc Adams, F. C. et al.} 1993
        {\it Phys.\ Rev.} {\bf D47}, 426.

\bibitem[Albrecht \& Steinhardt 1982]{as}{\sc Albrecht, A. \& Steinhardt, P.
J.}
    1982 {\it Phys.\ Rev.\ Lett.} {\bf 48}, 1220.

\bibitem[Alcock et al. 1995]{macho1}{\sc Alcock, C. et al.} 1995
        {\it Phys.\ Rev.\ Lett.} {\bf 74}, 2867.

\bibitem[Athanassopoulos et al. 1995]{numass2}{\sc Athanassopoulos, C. et al.}
1995
 {\it Phys. Rev. Lett.} {\bf 75}, 2650.

\bibitem[Babul \& Katz 1993]{outs}{\sc Babul, A. \& Katz, N.} 1993
        {\it Astrophys.\ J.} {\bf 406}, L51.

\bibitem[Bahcall et al. 1994]{nostars1}{\sc Bahcall, J. et al.} 1994
        {\it Astrophys.\ J.\ Lett.} {\bf 435}, L51.

\bibitem[Banks et al. 1995]{banks3}{\sc Banks, T. et al.} 1995 hep-th/9503114.

\bibitem[Bardeen, Steinhardt \& Turner 1983]{scalar4}{\sc Bardeen, J. M.,
    Steinhardt, P. J. \& Turner, M. S.} 1983 {\it Phys.\ Rev.\ D} {\bf 28},
697.

\bibitem[Bartlett et al. 1995]{lhccdm}{\sc Bartlett, J. et al.} 1995
        {\it Science} {\bf 267}, 980.

\bibitem[Becker-Szendy 1992]{numass6}{\sc Becker-Szendy, R. et al.}
        1992 {\it Phys.\ Rev.\ D} {\bf 46}, 3720.

\bibitem[Bennett et al. 1996]{dmr4yr1}{\sc Bennett, C. L. et al.} 1996
        {\it Astrophys.\ J.\ (Lett.)} in press, astro-ph/9601067.

\bibitem[Blumenthal et al. 1984]{faberetal}{\sc Blumenthal, G. et al.}
  1984 {\it Nature} {\bf 311}, 517.

\bibitem[Bolte \& Hogan 1995]{age2}{\sc Bolte, M. \& Hogan, C. J.} 1995 {\it
Nature}
        {\bf 376}, 399.

\bibitem[Bond \& Efstathiou, 1984]{boltzmann3}{\sc Bond, J. R. \& Efstathiou,
G.} 1984
        {\it Astrophys.\ J.} {\bf 285}, L45.

\bibitem[Bond \& Efstathiou 1991]{taucdm2}{\sc Bond, J. R. \& Efstathiou, G.}
1991
        {\it Phys.\ Lett.\ B} {\bf 265}, 245.

\bibitem[Briel et al. 1992]{gasratio2}{\sc Briel, U. G. et al.} 1992
  {\it Astron.\ Astrophys.} {\bf 259}, L31.

\bibitem[Brustein \& Veneziano 1994]{banks2}{\sc Brustein, R. \& Veneziano, G.}
    1994 {\it Phys.\ Lett.\ B} {\bf 329}, 429.

\bibitem[Bucher, Goldhaber, \& Turok 1995]{pu1}{\sc Bucher, M.,
	Goldhaber, A. S. \& Turok, N.} 1995 {\it Phys. Rev. D}
         {\bf 52}, 3314.

\bibitem[Bucher \& Turok 1995]{pu2}{\sc Bucher, M. \& Turok, N.} 1995
       hep-ph/9503393.

\bibitem[Burles \& Tytler 1996]{D5}{\sc Burles, S. \& Tytler, D.} 1996
astro-ph/9603070.

\bibitem[Caldwell 1995]{neutralinos}{\sc Caldwell, D. O.} 1995 in {\it Particle
and Nuclear
Astrophysics in the Next Millennium} (eds. E.W.~Kolb and
R.D.~Peccei), p.\ 38. World Scientific, Singapore.

\bibitem[Carswell et al. 1994]{D2}{\sc Carswell, R. F. et al.} 1994
        {\it Mon.\ Not.\ Roy.\ Astr.\ Soc.} {\bf 268}, L1.

\bibitem[Carswell, R. F. et al. 1996]{D7}{\sc Carswell, R. F. et al.} 1996
        {\it Mon.\ Not.\ Roy.\ Astron.\ Soc.} {\bf 278}, 506.

\bibitem[Cen et al. 1992]{tcdm1}{\sc Cen, R., Gnedin, N., Kofman, L., \&
        Ostriker, J. P.}  1992 {\it Astrophys.\ J.} {\bf 399}, L11.

\bibitem[Chaboyer et al. 1996]{age1}{\sc Chaboyer, B., Demarque, P.,
        Kernan, P. J. \& Krauss, L. M.} 1996 {\it Science} {\bf 271}, 957.

\bibitem[Cohen et al. 1992]{ewbaryo1}{\sc Cohen, A., Kaplan, D. \& Nelson, A.}
1992
    {\it Annu.\ Rev.\ Nucl.\ Part.\ Sci.} {\bf 43}, 27.

\bibitem[Collins \& Hawking 1973]{collins}{\sc Collins, C. B. \& Hawking, S.
W.}
           1973 {\it Astrophys. J.} {\bf 180}, 317.

\bibitem[Copeland et al. 1994]{liddle}{\sc Copeland, E. J. et al.}
	1994 {\it Phys.\ Rev.\ D} {\bf 49}, 6410.

\bibitem[Copi, Schramm \& Turner 1995]{bbn}{\sc Copi, C., Schramm, D. N.
  \& Turner, M. S.} 1995 {\it Science} {\bf 267}, 192.

\bibitem[Cowan, Thieleman \& Truran 1991]{agereview}{\sc Cowan, J., Thieleman,
F.
        \& Truran, J.}  1991 {\it Ann.\ Rev.\ Astron.\ Astrophys.} {\bf 29},
447.

\bibitem[Cowie \& Songaila 1996]{D9}{\sc Cowie, L. \& Songaila, A.} 1996 in
preparation.

\bibitem[Cvetic et al. 1990]{pati}{\sc Cvetic, M., Hubsch, T., Pati, J. \&
Stremnitzer, H.}
           1990 {\it Phys.\ Rev.\ D} {\bf 40}, 1311.

\bibitem[van Dalen \& Schaefer 1992]{boltzmann5}{\sc Dalen, A. van \& Schaefer,
R. K.} 1992
         {\it Astrophys.\ J.} {\bf 398}, 33.

\bibitem[Davis et al. 1992]{tcdm2}{\sc Davis, R. et al.} 1992 {\it Phys.\ Rev.\
Lett.}
        {\bf 69}, 1856.

\bibitem[Davis, Summers \& Schlegel 1992]{nucdm2}{\sc Davis, M., Summers,
        F. \& Schlegel, D.} 1992 {\it Nature} {\bf 359}.

\bibitem[Dekel 1994]{dekel}{\sc Dekel, A.} 1994 {\it Ann.\ Rev.\ Astron.\
Astrophys.}
{\bf 32}, 319.

\bibitem[Dodelson, Gates \& Stebbins 1996]{US}{\sc Dodelson, S., Gates, E.
        \& Stebbins, A. S.} 1996
        {\it Astrophys.\ J.}, in press, astro-ph/9509147.

\bibitem[Dodelson, Gates \& Turner 1996]{dgt}{\sc Dodelson, S., Gates, E. I. \&
Turner, M. S.}
   1996 astro-ph/9603081.

\bibitem[Dodelson, Gyuk \& Turner 1994]{taucdm1}{\sc Dodelson, S., Gyuk, G.
        \& Turner, M. S.} 1994
        {\it Phys.\ Rev.\ Lett.} {\bf 72}, 3578.

\bibitem[Dodelson \& Jubas 1993]{boltzmann7}{\sc Dodelson, S. \& Jubas, J. M.}
1993
        {\it Phys.\ Rev.\ Lett.} {\bf 70}, 2224.

\bibitem[Dolgov 1992]{ewbaryo2}{\sc Dolgov, A.} 1992 {\it Phys.\ Repts.} {\bf
222}, 309.

\bibitem[Efstathiou et al. 1990]{lcdm4}{\sc Efstathiou, G. et al.}
        1990 {\it Nature} {\bf 348}, 705.

\bibitem[Fabbri \& Pollock 1983]{tensor2}{\sc Fabbri, R. \& Pollock, M.} 1983
 {\it Phys.\ Lett.\ B} {\bf 125}, 445.

\bibitem[Flynn, Gould \& Bahcall 1996]{nostars2}{\sc Flynn, C., Gould, A. \&
Bahcall, J.}
 1996 astro-ph/9603035.

\bibitem[Freedman et al. 1994]{h_03}{\sc Freedman, W. et al.} 1994
        {\it Nature} {\bf 371}, 757.

\bibitem[Freese, Frieman, \& and Olinto 1990]{unnatural}{\sc Freese, K.,
Frieman, J. A.
    \& and Olinto, A.} 1990 {\it Phys.\ Rev.\ Lett.} {\bf 65}, 3233.

\bibitem[Fukuda et al. 1994]{numass5}{\sc Fukuda, Y. et al.} 1994
        {\it Phys.\ Lett.\ B} {\bf 335}, 237.

\bibitem[Fukugita, Hogan \& Peebles 1993]{h0rev1}{\sc Fukugita, M., Hogan, C.
J. \&
	Peebles, P. J. E.} 1993 {\it Nature} {\bf 366}, 309.

\bibitem[Fukugita, Hogan \& Peebles 1996]{fhp}{\sc Fukugita, M., Hogan, C. J.
\& Peebles, P. J. E.}
        1996 {\it Nature} {\bf 381}, 489.

\bibitem[Gasperini \& Veneziano 1994]{banks1}{\sc Gasperini, M. \& Veneziano,
G.}
       1994 {\it Phys.\ Rev.\ D} {\bf 50}, 2519.

\bibitem[Gates, Gyuk \& Turner 1995]{macho2}{\sc Gates, E., Gyuk, G.
        \& Turner, M. S.} 1995 {\it Phys.\ Rev.\ Lett.} {\bf 74}, 3724.

\bibitem[Gorski et al. 1996]{dmr4yr2}{\sc Gorski, K. M. et al.}
        1996 {\it Astrophys.\ J.\ (Lett.)} in press.

\bibitem[Guth 1981]{guth}{\sc Guth, A. H.} 1981 {\it Phys.\ Rev.\ D} {\bf 23},
347.

\bibitem[Guth \& Pi 1982]{scalar1}{\sc Guth, A. H. \& Pi, S.-Y.} 1982 {\it
Phys.\ Rev.\ Lett.}
{\bf 49}, 1110.

\bibitem[Hamuy 1995]{h_02}{\sc Hamuy, M. et al.} 1995 {\it Astron.\ J.} {\bf
109}, 1.

\bibitem[Hawking 1982]{scalar2}{\sc Hawking, S. W.} 1982 {\it Phys.\ Lett.\ B}
{\bf 115}, 295.

\bibitem[Hill 1995]{numass3}{\sc Hill, J. E.} 1995 {\it Phys.\ Rev.\ Lett.},
2654.

\bibitem[Hirata et al. 1992]{numass4}{\sc Hirata, K. S. et al.} 1992
   {\it Phys.\ Lett.\ B} {\bf 280}, 146.

\bibitem[Holman, Ramond \& Ross 1984]{florida}{\sc Holman, R., Ramond, P. \&
Ross, G. G.}
           1984 {\it Phys.\ Lett.\ B} {\bf 137}, 343.

\bibitem[Holtzman 1989]{boltzmann4}{\sc Holtzman, J. A.} 1989 {\it Astrophys.\
J.\ Suppl.}
           {\bf 71}, 1.

\bibitem[Hu \& Sugiyama 1995]{cbranisotropy}{\sc Hu, W. \& Sugiyama, N.} 1995
        {\it Phys.\ Rev.\ D} {\bf 51}, 2599.

\bibitem[Hu \& White 1996]{omegacmb3}{\sc Hu, W. \& White, M.} 1996
astro-ph/9602020.

\bibitem[Hu, Turner \& Weinberg 1994]{htw}{\sc Hu, Y., Turner, M. S. \&
Weinberg, E. J.} 1994
{\it Phys.\ Rev.\ D} {\bf 49}, 3830.

\bibitem[Huchra \& Geller 1989]{structure}{\sc Huchra, J. \& Geller, M.} 1989
{\it Science}
 {\bf 246}, 897.

\bibitem[Jacoby G. et al. 1992]{h0rev2}{\sc Jacoby, G. et al.} 1992
   {\it Proc. Astron.\ Soc.\ Pac.} {\bf 104}, 599.

\bibitem[Jensen \& Schabes 1987]{nohair2}{\sc Jensen, L. \& Stein-Schabes, J.}
1987
  {\it Phys.\ Rev.\ D} {\bf 35}, 1146.

\bibitem[Jungman et al. 1996a]{omegacmb2}{\sc Jungman, G., Kamionkowski, M.,
        Kosowsky, A. \& Spergel, D.} 1996a
        {\it Phys.\ Rev.\ Lett.} {\bf 76}, 1007.

\bibitem[Jungman et al. 1996b]{learn2}{\sc Jungman, G., Kamionkowski, M.,
        Kosowsky, A. \& Spergel, D.} 1996b
        {\it Phys.\ Rev.\ D}, in press, astro-ph/9512139.

\bibitem[Kamionkowski, Spergel \& Sugiyama 1994]{omegacmb1}{\sc Kamionkowski,
M.,
        Spergel, D. N. \& Sugiyama, N.} 1994
        {\it Astrophys. J.} {\bf 426}, L57.

\bibitem[Knox 1995]{learn1}{\sc Knox, L.} 1995 {\it Phys.\ Rev.\ D} {\bf 52},
4307.

\bibitem[Knox \& Turner 1993]{knox}{\sc Knox, L. \& Turner, M. S.} 1993
    {\it Phys.\ Rev.\ Lett.} {\bf 70}, 371.

\bibitem[Knox, L. \& Turner 1994]{knoxmst}{\sc Knox, L. \& Turner, M. S.}
	1994 {\it Phys. Rev. Lett.} {\bf 73}, 3347.

\bibitem[Kochanek 1996]{qsolensing}{\sc Kochanek, C. S.} 1996 {\it Astrophys.\
J.},
in press, astro-ph/9510077.

\bibitem[Kofman, Linde \& Einsato 1987]{multiple2}{\sc Kofman, L. A., Linde, A.
D.
        \& Einsato, J.} 1987 {\it Nature} {\bf 326}, 48.

\bibitem[Kofman \& Starobinsky 1985]{lcdm5}{\sc Kofman, L. \& Starobinsky, A.
A.} 1985
        {\it Sov.\ Astron.\ Lett.} {\bf 11}, 271.

\bibitem[Kolb 1991]{kolbreview}{\sc Kolb, E. W.} 1991 {\it Physica\ Scripta}
           {\bf T36}, 199.

\bibitem[Kolb \& Turner 1983]{kt}{\sc Kolb, E. W. \& Turner, M. S.} 1983 {\it
Ann.\ Rev.\ Nucl.\ Part.\ Sci.} {\bf 33}, 645.

\bibitem[Kolb \& Turner 1990]{eu}{\sc Kolb, E. W. \& Turner, M. S.} 1990 {\it
The Early
Universe}. Addison-Wesley, Redwood City.

\bibitem[Kosowsky, Turner, \& Watkins 1992]{vacuumpop2}{\sc Kosowsky, A.,
Turner, M. S. \& Watkins, R.}
   1992 {\it Phys. Rev. Lett.} {\bf 69}, 2026.

\bibitem[Krauss \& Turner 1995]{bestfit1}{\sc Krauss, L. \& Turner, M. S.}
        1995 {\it Gen.\ Rel.\ Grav.} {\bf 27},
        1137.

\bibitem[La \& Steinhardt 1991]{lapjs}{\sc La, D. \& Steinhardt, P. J.} 1991
    {\it Phys.\ Rev.\ Lett.} {\bf 62}, 376.

\bibitem[Lanzetta, Wolfe \& Turnshek 1995]{Dlya1}{\sc Lanzetta, K., Wolfe, A.
M.
        and Turnshek, D. A.} 1995
        {\it Astrophys.\ J.} {\bf 440}, 435.

\bibitem[Liddle et al. 1995]{nucontent2}{\sc Liddle, A. et al.} 1995
astro-ph/9511957.

\bibitem[Liddle \& Lyth 1992]{tcdm5}{\sc Liddle, A. \& Lyth, D.} 1992
        {\it Phys. Lett. B} {\bf 291}, 391.

\bibitem[Liddle \& Lyth 1993]{jpo-ll2}{\sc Liddle, A. \& Lyth, D.} 1993
        {\it Phys.\ Repts.} {\bf 231}, 1.

\bibitem[Lidsey \& Coles 1992]{tcdm6}{\sc Lidsey, J. E. \& Coles, P.} 1992
        {\it Mon.\ Not.\ Roy.\ Astron.\ Soc.} {\bf 258}, 57.

\bibitem[Lidsey et al. 1995]{reconstruct2}{\sc Lidsey, J. et al.} 1995
astro-ph/9508078.

\bibitem[Lin 1996]{biasing}{\sc Lin, H. et al.} 1996 {\it Astrophys.\ J.}, in
press,
     astro-ph/9602064.

\bibitem[Linde 1982]{linde}{\sc Linde, A. D.} 1982 {\it Phys.\ Lett.\ B} {\bf
108}, 389.

\bibitem[Linde 1983]{chaotic}{\sc Linde, A. D.} 1983 {\it Phys.\ Lett.\ B} {\bf
129}, 177.

\bibitem[Linde 1990]{eternal}{\sc Linde, A. D.}, 1990 {\it Inflation\ and\
Quantum\ Cosmology}.
      Academic Press.

\bibitem[Lucchin, Mattarese \& Mollerach 1992]{tcdm3}{\sc Lucchin, F.,
Mattarese, S., \&
        Mollerach, S.} 1992 {\it Astrophys.\ J.} {\bf 401}, L49.

\bibitem[Ma \& Bertschinger 1994]{nucontent1}{\sc Ma, C. P. \& Bertschinger,
E.} 1994
{\it Astrophys.\ J.} {\bf 434}, L5.

\bibitem[Murayama 1994]{lbl}{\sc Murayama, H. et al.} 1994 {\it Phys.\ Rev.\ D}
{\bf 50}, R2356.

\bibitem[Novikov \& Zel'dovich 1967]{araa1}{\sc Novikov, I. D. \&
  Zel'dovich, Ya. B.} 1967 {\it Ann.\ Rev.\
 Astron.\ Astrophys.} {\bf 5}, 627.

\bibitem[Novikov \& Zel'dovich 1973]{araa2}{\sc Novikov, I. D. \&
  Zel'dovich, Ya. B.} 1973 {\it Ann.\ Rev.\
 Astron.\ Astrophys.} {\bf 11}, 387.

\bibitem[Olive 1990]{olive}{\sc Olive, K.} 1990 {\it Phys.\ Repts.} {\bf 190},
309.

\bibitem[Ostriker 1993]{jpo-ll1}{\sc Ostriker, J. P.} 1993 {\it Ann.\ Rev.\
Astron.\
        Astrophys.} {\bf 31}, 689.

\bibitem[Ostriker \& Steinhardt 1995]{bestfit2}{\sc Ostriker, J. P. \&
Steinhardt, P. J.}
	1995 {\it Nature} {\bf 377}, 600.

\bibitem[Parke 1995]{numass1}{\sc Parke, S.} 1995 {\it Phys.\ Rev.\ Lett.}
        {\bf 74}, 839.

\bibitem[Peacock \& Dodds 1994]{pd}{\sc Peacock, J. \& Dodds, S.} 1994
        {\it Mon.\ Not.\ Roy.\ Astron.\ Soc.} {\bf 267}, 1020.

\bibitem[Peebles 1984]{lcdm3}{\sc Peebles, P. J. E.} 1984 {\it Astrophys.\ J.}
{\bf 284}, 439.

\bibitem[Peebles et al. 1991]{bigbang1a}{\sc Peebles, P. J. E., Schramm, D. N.,
Turner, E., \& Kron, R.} 1991 {\it Nature} {\bf 352}, 769.

\bibitem[Peebles \& Yu 1970]{boltzmann1}{\sc Peebles P. J. E. \& Yu, J. T.}
1970
        {\it Astrophys.\ J.} {\bf 162}, 815.

\bibitem[Perlmutter et al. 1996]{lblq_0}{\sc Perlmutter, S. et al.} 1996
astro-ph/9602122.

\bibitem[Pi 1984]{pi2}{\sc Pi, S.-Y.} 1984 {\it Phys.\ Rev.\ Lett.} {\bf 52},
1725.

\bibitem[Pogosyan \& Starobinsky 1995]{nucdm4}{\sc Pogosyan, D. \& Starobinsky,
A. A.}
        1995 astro-ph/9502019.

\bibitem[Preskill 1984]{preskill}{\sc Preskill, J.} 1984 {\it Ann.\ Rev.\
Nucl.\ Part.\ Sci.}
{\bf 34}, 461.

\bibitem[Primack et al. 1995]{nucdm3}{\sc Primack, J. et al.} 1995
        {\it Phys.\ Rev.\ Lett.} {\bf 74}, 2160.

\bibitem[Randall, Soljacic \& Guth 1996]{lisa}{\sc Randall, L., Soljacic, M.
   \& Guth, A. H.} 1996 hep-ph/9601296.

\bibitem[Ratra 1992]{bfield2}{\sc Ratra, B.} 1992 {\it Astrophys. J.} {\bf
391}, L1.

\bibitem[Reiss, Kirshner \& Press 1995]{h_01}{\sc Reiss, A., Kirshner, R. P.
        \& Press, W.} 1995 {\it Astrophys.\ J.} {\bf 438}, L17.

\bibitem[Rosenberg 1995]{llnl}{\sc Rosenberg, L. J.} 1995
        {\it Particle\ World} {\bf 4}, 3.

\bibitem[Rubakov, Sazhin \& Veryaskin 1984]{tensor1}{\sc Rubakov, V. A.,
          Sazhin, M. \& Veryaskin, A.} 1982 {\it Phys.\ Lett.\ B}
          {\bf 115}, 189.

\bibitem[Rugers \& Hogan 1996a]{D3}{\sc Rugers, M. \& Hogan, C. J.} 1996a
        {\it Astrophys.\ J.\ (Lett.)}, in press, astro-ph/9512004.

\bibitem[Rugers \& Hogan 1996b]{D8}{\sc Rugers, M. \& Hogan, C. J.}
	1996b astro-ph/9603084.

\bibitem[Salopek 1992]{tcdm4}{\sc Salopek, D.} 1992
	{\it Phys.\ Rev.\ Lett.} {\bf 69}, 3602.

\bibitem[Shafi \& Stecker 1984]{nucdm1}{\sc Shafi, Q. \& Stecker, F.} 1984
        {\it Phys.\ Rev.\ Lett.} {\bf 53}, 1292.

\bibitem[Shafi \& Vilenkin 1984]{pi1}{\sc Shafi, Q. \& Vilenkin, A.} 1984
  {\it Phys.\ Rev.\ Lett.} {\bf 52}, 691.

\bibitem[Silk \& Turner 1986]{multiple1}{\sc Silk, J. \& Turner, M. S.} 1986
   {\it Phys.\ Rev.\ D} {\bf 35}, 419.

\bibitem[Smoot et al. 1992]{dmr}{\sc Smoot, G. F. et al.} 1992
        {\it Astrophys. J.} {\bf 396}, L1.

\bibitem[Songaila et al. 1994]{D1}{\sc Songaila, A. et al.} 1994 {\it Nature}
{\bf 368}, 599.

\bibitem[Souradeep \& Sahni 1992]{tcdm7}{\sc Souradeep, T. \& Sahni, V.} 1992
        {\it Mod.\ Phys.\ Lett.\ A} {\bf 7}, 3541.

\bibitem[Starobinsky 1982]{scalar3}{\sc Starobinsky, A. A.} 1982 {\it Phys.\
Lett.\ B}
        {\bf 117}, 175.

\bibitem[Starobinsky 1983a]{tensor3}{\sc Starobinsky, A. A.} 1983a
 {\it Sov.\ Astron.\ Lett.} {\bf 9}, 302.

\bibitem[Starobinsky 1983b]{nohair3}{\sc Starobinsky, A. A.} 1983b {\it JETP\
Lett.}
      {\bf 37}, 66.

\bibitem[Steinhardt \& Turner 1984]{pjsmst}{\sc Steinhardt, P. J. \& Turner, M.
S.}
    1984 {\it Phys.\ Rev.\ D} {\bf 29}, 2162.

\bibitem[Storrie-Lombardi et al. 1995]{Dlya2}{\sc Storrie-Lombardi, L. J.,
        McMahon, R. G., Irwin, M. J. \& Hazard, C.} 1995
        astro-ph/9503089.

\bibitem[Strauss \& Willick 1995]{strauss}{\sc Strauss, M. \& Willick, J.} 1995
        {\it Phys.\ Repts.} {\bf 261}, 271.

\bibitem[Sugiyama \& Gouda 1992]{boltzmann6}{\sc Sugiyama, N. \& Gouda, N.}
1992
        {\it Prog.\ Theor.\ Phys.} {\bf 88}, 803.

\bibitem[Trimble 1987]{darkmatter1}{\sc Trimble, V.} 1987 {\it Ann.\ Rev.\
Astron.\
Astrophys.} {\bf 25}, 425.

\bibitem[Turner 1987]{darkmatter2}{\sc Turner, M. S.} 1993 {\it Proc.\ Nat.\
Acad.\ Sci.\ (USA)} {\bf 90}, 4827.

\bibitem[Turner 1991]{pdm}{\sc Turner, M. S.} 1991 {\it Physica Scripta}
            {\bf T36}, 167.

\bibitem[Turner 1993a]{bigbang1b}{\sc Turner, M. S.} 1993a {\it Science} {\bf
262}, 861.

\bibitem[Turner 1993b]{reconstruct1}{\sc Turner, M. S.} 1993b {\it Phys.\ Rev.\
D} {\bf 48}, 3502.

\bibitem[Turner 1996]{tlw2}{\sc Turner, M. S.} 1996 astro-ph/9607066.

\bibitem[Turner, Lidsey, \& White 1993]{tlw1}{\sc Turner, M. S., Lidsey, J. \&
White, M.}
		1993 {\it Phys.\ Rev.\ D}
{\bf 48}, 4613.

\bibitem[Turner, Steigman \& Krauss 1984]{lcdm1}{\sc Turner, M. S., Steigman,
G.
	\& Krauss, L.} 1984 {\it Phys.\ Rev.\ Lett.} {\bf 52}, 2090.

\bibitem[Turner \& Widrow 1986]{nohair1}{\sc Turner, M. S. \& Widrow, L. M.}
        1986 {\it Phys.\ Rev.\ Lett.} {\bf 57}, 2237.

\bibitem[Turner \& Widrow 1988]{bfield1}{\sc Turner, M. S. \& Widrow, L. M.}
1988
    {\it Phys. Rev. D} {\bf 37}, 2743.

\bibitem[Turner \& Wilczek 1990]{vacuumpop1}{\sc Turner, M. S. \& Wilczek, F.}
  1990 {\it Phys.\ Rev.\ Lett.} {\bf 65}, 3080.

\bibitem[Tytler, Fan \& Burles 1996]{D4}{\sc Tytler, D., Fan, K.-M.
        \& Burles, S.} 1996 astro-ph/9603069.

\bibitem[Wampler et al. 1996]{D6}{\sc Wampler, E. J. et al.} 1996
         {\it Astron.\ Astrophys.}, in press, astro-ph/9512084.

\bibitem[Weinberg 1989]{cosmoconst}{\sc Weinberg, S.} 1989
        {\it Rev.\ Mod.\ Phys.} {\bf 61}, 1.

\bibitem[White \& Fabian 1995]{gasratio3}{\sc White, D. A. \& Fabian, A. C.}
        1995 {\it Mon.\ Not.\ Roy.\ Astron.\ Soc.} {\bf 273}, 72.

\bibitem[White \& Bunn 1995]{dmr4yr3}{\sc White, M. \& Bunn, E. F.} 1995
        {\it Astrophys.\ J.\ (Lett.)} {\bf 450}, 477.

\bibitem[White, Scott, \& and Silk 1995]{cbrsummary}{\sc White, M., Scott, D.
         \& Silk, J.} 1995 {\it Science} {\bf 268}, 829.

\bibitem[White et al. 1995]{stp}{\sc White, M., Scott, D.,
        Silk, J. \& Davis, M.} 1995 {\it Mon.\ Not.\ Roy.\ Astron.\ Soc.} {\bf
276}, L69.

\bibitem[White et al. 1993]{gasratio1}{\sc White, S. D. M. et al.}
     1993 {\it Nature} {\bf 366}, 429.

\bibitem[White, Efstathiou \& Frenk 1993]{sigma8cluster}{\sc White, S. D. M.,
Efstathiou, G.
        \& Frenk, C. S.} 1993
        {\it Mon.\ Not.\ Roy.\ Astron.\ Soc.} {\bf 262}, 1023.

\bibitem[White, Frenk \& Davis 1983]{hdm}{\sc White, S. D. M., Frenk, C.
   \& Davis, M.} 1983 {\it Astrophys.\ J.} {\bf 274}, L1.

\bibitem[Wilson \& Silk 1981]{boltzmann2}{\sc Wilson, M. L. \& Silk, J.} 1981
        {\it Astrophys.\ J.} {\bf 243}, 14.

\bibitem[Zel'dovich \& Novikov 1983]{relastro}{\sc Zel'dovich, Ya. B.
 \& Novikov, I. D.} 1983 {\it Relativistic
Astrophysics (Structure and Evolution of the Universe)}. Univ. Chicago
Press.


\end{thebibliography}
\end{document}